\documentclass[aps,prl,twocolumn,superscriptaddress,notitlepage,nofootinbib,longbibliography]{revtex4-1}
\usepackage{mathrsfs,color}
\usepackage{booktabs}
\usepackage{float}
\usepackage{tabularx}
\usepackage{epsfig,float}
\usepackage{graphicx}
\usepackage{amsfonts}
\usepackage[figuresright]{rotating}
\usepackage{amssymb}
\usepackage{amsmath}
\usepackage{dcolumn}
\usepackage{bm}
\usepackage{comment}
\usepackage{xcolor}
\usepackage{color}
\usepackage{braket}
\usepackage{units}
\usepackage{xspace}

\usepackage{bbold}
\definecolor{mypink3}{cmyk}{0, 0.7808, 0.4429, 0.1412}
\definecolor{mypink1}{rgb}{0.858, 0.188, 0.478}
\definecolor{mypink2}{RGB}{219, 48, 122}
\usepackage[colorlinks=true, allcolors=mypink1]{hyperref}
\usepackage[fleqn]{mathtools}

\usepackage[shortlabels]{enumitem}

\begin{document}
	
\title{Experimental Observation of Dynamical Phase Transitions in a Dephased Photonic Quantum Walk}
	
\author{Xiaojian Huang}\thanks{These authors contributed equally}
\affiliation{Beijing Computational Science Research Center, Beijing 100193, China}
\author{Lei Xiao}\thanks{These authors contributed equally}
\affiliation{School of Physics, Southeast University, Nanjing 211189, China}
\author{Bingzi Huo}
\affiliation{Beijing Computational Science Research Center, Beijing 100193, China}
\author{Xiaowei Wang}
\affiliation{Beijing Computational Science Research Center, Beijing 100193, China}
\author{Stefano Longhi}
\affiliation{Dipartimento di Fisica, Politecnico di Milano, Piazza Leonardo da Vinci 32, I-20133 Milano, Italy}
\affiliation{IFISC (UIB-CSIC), Instituto de Fisica Interdisciplinar y Sistemas Complejos, E-07122 Palma de Mallorca, Spain}
\author{Peng Xue}
\email{gnep.eux@gmail.com}
\affiliation{School of Physics, Southeast University, Nanjing 211189, China}

\begin{abstract}
Dynamical phase transitions in open quantum systems govern how non-equilibrium states relax toward a stationary state. We study these transitions experimentally using a discrete-time photonic quantum walk on a three-node graph. A tunable synthetic gauge flux and calibrated dephasing allow us to control time-reversal symmetry and the detailed balance properties of the effective Markovian dynamics. With detailed balance, we observe a first-order dynamical phase transition marked by a crossing of real Liouvillian eigenvalues. When detailed balance is broken, we observe a second-order dynamical phase transition at an exceptional point where eigenvalues and eigenvectors coalesce. By progressively reducing the dephasing strength, we track the crossover toward the quantum-coherent regime and determine that the transitions persist down to a finite threshold. Our results link Liouvillian spectral topology to relaxation criticality and demonstrate a controllable platform for engineered dissipative dynamics.
\end{abstract}

\maketitle
\emph{Introduction}.
Dynamical phase transitions have been widely explored in closed quantum systems, often through quantities such as the Loschmidt echo~\cite{Heyl_Review_2018,Heyl_PRL_2013,Wang_PRL_2019,Marino_2022}. Realistic physical platforms, however, are inherently open, and extending these ideas to systems coupled to an environment remains challenging. In this open-system setting, approaching the thermodynamic limit is a necessary condition for the observation of a dissipative phase transition \cite{Fitzpatrick2017,Rodriguez2017,Fink2018,minganti2018spectral}, while the mechanisms underlying dynamical critical behavior remain poorly understood~\cite{Gao_PRL_2024,Zunkovic_PRL_2018}.

Recent theoretical works have connected dynamical phase transitions to the spectral properties of the generator of the dynamics. In classical Markov processes, Teza \textit{et al.}~\cite{Teza_PRL_2023} showed that {\em the relaxation dynamics} can display phase-transition-like behavior without the need for the Liouvillian spectral gap to close~\cite{mori2023symmetrized,minganti2018spectral,casteels2017critical}.
Unlike equilibrium phase transitions -- typically characterized by scaling laws and thermodynamic critical exponents -- these transitions arise from qualitative changes in the spectral structure of the dynamics. They appear as eigenvalue crossings of classical Markov generators, leading to abrupt rearrangements of relaxation modes, rather than non-equilibrium steady states.
In systems obeying detailed balance, time-reversal symmetry (TRS) constrains the spectrum to be real, allowing only first-order dynamical phase transitions (FOPTs) via eigenvalue crossings~\cite{Teza_PRL_2023,Blythe_JPCS_2006}. Second-order dynamical phase transitions (SOPTs), associated with the coalescence of eigenvalues and eigenvectors at exceptional points (EPs)~\cite{xiao2024non,shi2025enhanced,arkhipov2021generating,xue2020non}, are therefore forbidden~\cite{lamb1998time,choi2023noninvertible,sato2012time,parzygnat2023time}. Their emergence requires the spectrum of the Markov generator to become complex~\cite{uff1,uff2,uff3,Longhi_AQT_2025}. One route is to break detailed balance by introducing a synthetic gauge flux, which breaks TRS and enables SOPTs driven by non-Hermitian degeneracies in the hybrid quantum--classical regime~\cite{Longhi_AQT_2025,BH2024,uffa4}.
Despite this theoretical progress, realizing these ideas in a controlled quantum platform poses several challenges.

Open quantum walks (OQWs) have emerged as a powerful and widely used framework for describing discrete-time open quantum dynamics~\cite{Kwiat_Science_2000,Kendon_PRA_2003,
				Brun_PRA_2003,Kendon_LNP_2004,Kendon_MSCS_2007,broome2010discrete,
				Annabestani_PRA_2010, Schreiber_PRL_2011,Chalabi_PRL_2019,VenegasAndraca_QIP_2012,Zhang_CPB_2013,
				Alberti_NJP_2014,Caruso_NC_2016}, where both coherent evolution and dissipation are encoded in the spectrum of a quantum map rather than in a continuous-time Lindblad generator. In particular, photonic OQWs offer several advantages, providing a feasible platform for the observation of dynamical phase transitions and the role of TRS. First, they enable interferometric phase coherence preservation to imprint a synthetic flux while independently tuning dephasing over a wide range~\cite{broome2010discrete,Schreiber_PRL_2011,Chalabi_PRL_2019}.
Second, they allow access to the spectrum of the quantum generator through reconstruction of the quantum channel and its eigenmodes via quantum process tomography~\cite{Chuang_Nielsen_1997,Poyatos_PRL_1997,OBrien_PRL_2004,Samach_PRApplied_2022}.
Third, EP-driven criticality and eigenvector coalescence remain robust under realistic decoherence and parameter noise~\cite{Zhang_PRL_2019_EPNoise,Wiersig_PRA_2020}.

In this Letter, we report on the first experimental observation of both first- and second-order dynamical phase transitions in a photonic OQW. We further demonstrate that second-order transitions originate from EP degeneracies that are intrinsically tied to TRS breaking and stroboscopic dynamics, disappearing in the continuous-time (Lindbladian) dynamics. More broadly, our setup provides a versatile platform for engineering dissipative dynamics and exploring non-Hermitian phenomena in open quantum systems, with implications for non-Hermitian spectral topology~\cite{Bergholtz_RMP_2021,Okuma_ARCMP_2023,xiao2025non,DFM22,Xue_PRL_2026} and relaxation engineering such as Mpemba effects~\cite{uff2,uff3,Ares_NRP_2025,Joshi_PRL_2024}.

\emph{Theoretical framework}. We consider a discrete-time OQW on a three-node graph, governed by the interplay between coherent unitary evolution and dephasing. The unitary evolution $\mathcal{U}=e^{-i\mathrm{H}\tau}$ is governed by the Hamiltonian
\begin{equation}
\mathrm{H}=J_0|0\rangle\langle 1|+J_1|1\rangle\langle 2|+J_2 e^{i \phi}|0\rangle\langle 2|+\text{H.c.}
\end{equation}
with hopping amplitudes $J_0,J_1,J_2$ and a synthetic gauge flux $\phi$ threading the triangular loop.
We set \(J_0=J_1=1\) and \(J_2=0.5\), following the minimal three-node Floquet-walk model of Ref.~\cite{Longhi_AQT_2025}.
		This choice places the model in a symmetry-controlled regime, with a TRS-protected diabolic crossing at \(\phi=0\) and an EP transition when the synthetic flux breaks TRS.
		The dimensionless control parameter is \(\beta\equiv J_0\tau\), so that
		\(\mathcal{U}(\beta,\phi)=\exp[-i\mathrm{H}(\phi)\tau]\).
		Here \(\beta\) tunes the coherent step strength, while \(\phi\) controls the synthetic gauge flux and hence TRS.
 Decoherence is introduced by local pure dephasing in the node basis $|n \rangle$  \cite{Kendon_LNP_2004}, with Kraus operators of the dephasing channel being $\mathcal{K}_n=\ket{n}\bra{n}$ ($n=0,1,2$). They implement an unread which-node measurement in the graph-node basis after the coherent step: after letting
		\(\sigma=\mathcal{U}\rho^{(s)}\mathcal{U}^\dagger\), the map
		\(\sum_n\mathcal{K}_n\sigma\mathcal{K}_n^\dagger\) preserves node populations while removing inter-node coherences. The system dynamics is described by the following map for the density matrix (see e.g. \cite{broome2010discrete})
\begin{equation}
	\rho^{(s+1)}=(1-q)\,\mathcal{U}\rho^{(s)}\mathcal{U}^{\dagger}
	+q \sum_{n=0}^{2} \mathcal{K}_n\, \mathcal{U}\rho^{(s)}\mathcal{U}^{\dagger}\mathcal{K}_n^{\dagger} \equiv e^{-\mathcal{L}} \rho^{(s)},
	\label{eqMAr}
\end{equation}
where $s=0,1,2,\dots$ labels the discrete step, $q\in[0,1]$ controls the dephasing strength, and \( - \mathcal{L}\) is the effective Floquet--Liouvillian superoperator. This standard discrete-time quantum-walk channel interpolates between coherent quantum evolution ($q=0$) and a classicalized Markov process ($q=1$)~\cite{Kwiat_Science_2000,Kendon_PRA_2003,
				Brun_PRA_2003,Kendon_LNP_2004,Kendon_MSCS_2007,broome2010discrete,
				Annabestani_PRA_2010,VenegasAndraca_QIP_2012,Zhang_CPB_2013,
				Alberti_NJP_2014,Caruso_NC_2016}.
			Equation~(\ref{eqMAr}) defines the finite-step Floquet channel and should not be regarded as a Trotter discretization of an underlying continuous-time Lindblad equation. In fact, as demonstrated in the Supplemental Material~\cite{sm}, the continuous-time (Lindbladian) limit of Eq.~(\ref{eqMAr}), obtained by taking the double limit $\tau,q \rightarrow 0$ with $\gamma=q/\tau$ held constant, preserves time-reversal symmetry. Consequently, SOPTs cannot arise in this limit.
			
We first consider the fully dephased limit \(q=1\) as a transparent benchmark that isolates the transition mechanism and highlights the role of TRS breaking \cite{Longhi_AQT_2025}. In the fully dephased limit ($q=1$), all coherences are erased after each step, so the dynamics reduces to the population vector
$P^{(s)}\equiv\big(P_0^{(s)},P_1^{(s)},P_2^{(s)}\big)^{T}$ with $P_i^{(s)}\equiv\rho^{(s)}_{ii}$.
The evolution then follows a discrete-time Markov process given by
\begin{equation}
	P^{(s+1)}=Q\,P^{(s)}, \qquad Q_{nm}=|\mathcal{U}_{nm}|^2,
\end{equation}
where $n,m=0,1,2$ label the graph nodes.
Unitarity of $\mathcal{U}$ makes $Q$ doubly stochastic \cite{doublyS}, yielding the uniform stationary distribution
$P^{\rm ss}\equiv(1/3,1/3,1/3)^{T}$.
The relaxation modes \(r^{(k)}\) are the right eigenvectors of \(Q\),
\[
Q r^{(k)}=E_k r^{(k)},\qquad
\lambda_k=-\mathrm{Log}(E_k),
\]
with \(\lambda_k\) the corresponding Floquet exponents and \(\mathrm{Log}\) taken on the principal branch.
Away from the non-stationary degeneracy, we order the modes by their decay rates as	\[
0=\mathrm{Re}[\lambda_0]<\mathrm{Re}[\lambda_1]<\mathrm{Re}[\lambda_2].
	\]
Thus \(r^{(0)}=P^{\rm ss}\) is the stationary mode, while \(r^{(1)}\) and \(r^{(2)}\) are the slow and fast relaxation modes, respectively.
At the degeneracy, the two non-stationary decay rates become equal; as \(\beta\) is tuned through this point, the branch with the smaller decay rate is flipped, which is the defining spectral signature of a dynamical phase transition. These modes are collective eigenmodes of \(Q\), not node-basis states; the full spectral decomposition is given in Supplemental Material~\cite{sm}.

Under TRS (\(\phi=0\)), \(|\mathcal{U}_{ij}|^2=|\mathcal{U}_{ji}|^2\), so \(Q_{ij}=Q_{ji}\) and detailed balance holds.
Hence \(Q\) is real symmetric and diagonalizable, the eigenvalues are real, and any eigenvalue crossing must be diabolic. This means that under TRS the slow-relaxation branch switches abruptly as the control parameter is varied across the diabolic point, yielding a FOPT.
The diabolic crossing is not produced by dephasing alone: dephasing defines the Markov map \(Q_{nm}=|\mathcal{U}_{nm}|^2\), while \(\mathcal{U}(\beta,0)\) supplies the \(\beta\)-dependent transition probabilities whose spectrum crosses.
				
Breaking TRS (\(\phi\neq0\)) violates detailed balance and generically yields complex spectra. The two non-stationary relaxation modes can then coalesce at an EP and split into a complex-conjugate pair, producing damped oscillatory relaxation and a continuous relaxation-pattern change characteristic of a SOPT \cite{Longhi_AQT_2025}.
						
\emph{Experimental implementation}.
We implement the OQW in a phase-stable single-photon interferometric network, with the experimental setup shown in End Matter, Fig.~\ref{fig:EM_setup}.
Heralded single photons generated via type-II spontaneous parametric down-conversion are injected into the network, with one photon used as a trigger.
The qutrit basis is encoded in the polarization ($H,V$) and spatial ($U,D$) modes of a single photon,
$|0\rangle \Leftrightarrow |UH\rangle$, $|1\rangle \Leftrightarrow |UV\rangle$, and $|2\rangle \Leftrightarrow |DH\rangle$.
The platform comprises three modules: state preparation, interferometric evolution, and detection.

For each \((\beta,\phi)\), the target unitary \(\mathcal{U}(\beta,\phi)\) is synthesized with beam displacers (BDs) and wave plates, followed by a calibrated dephasing block that introduces controlled which-node distinguishability.
	The fully and partially dephased regimes use complementary modules, and \(q\) is calibrated independently from single-photon visibility; details are given in End Matter and Supplemental Material~\cite{sm}.
After each step, a BD separates the three output modes and avalanche photodiodes (APDs) record the counts $N_i(s)$, yielding
$P_i^{(s)}=N_i(s)/\sum_{m=0}^{2}N_m(s)$.
In the fully dephased benchmark ($q=1$), the relevant one-step map reduces to the population transfer matrix $Q$, which we reconstruct column by column by preparing the basis inputs $\ket{j}$ and measuring the output probabilities, $Q_{ij}=P_i^{(j)}$. From the reconstructed \(Q\), we extract the eigenvalues and Floquet exponents using the convention defined above.
	For \(0<q<1\), we reconstruct the full one-step superoperator.
\begin{figure}[htbp]
	\centering
	\includegraphics[width=\linewidth]{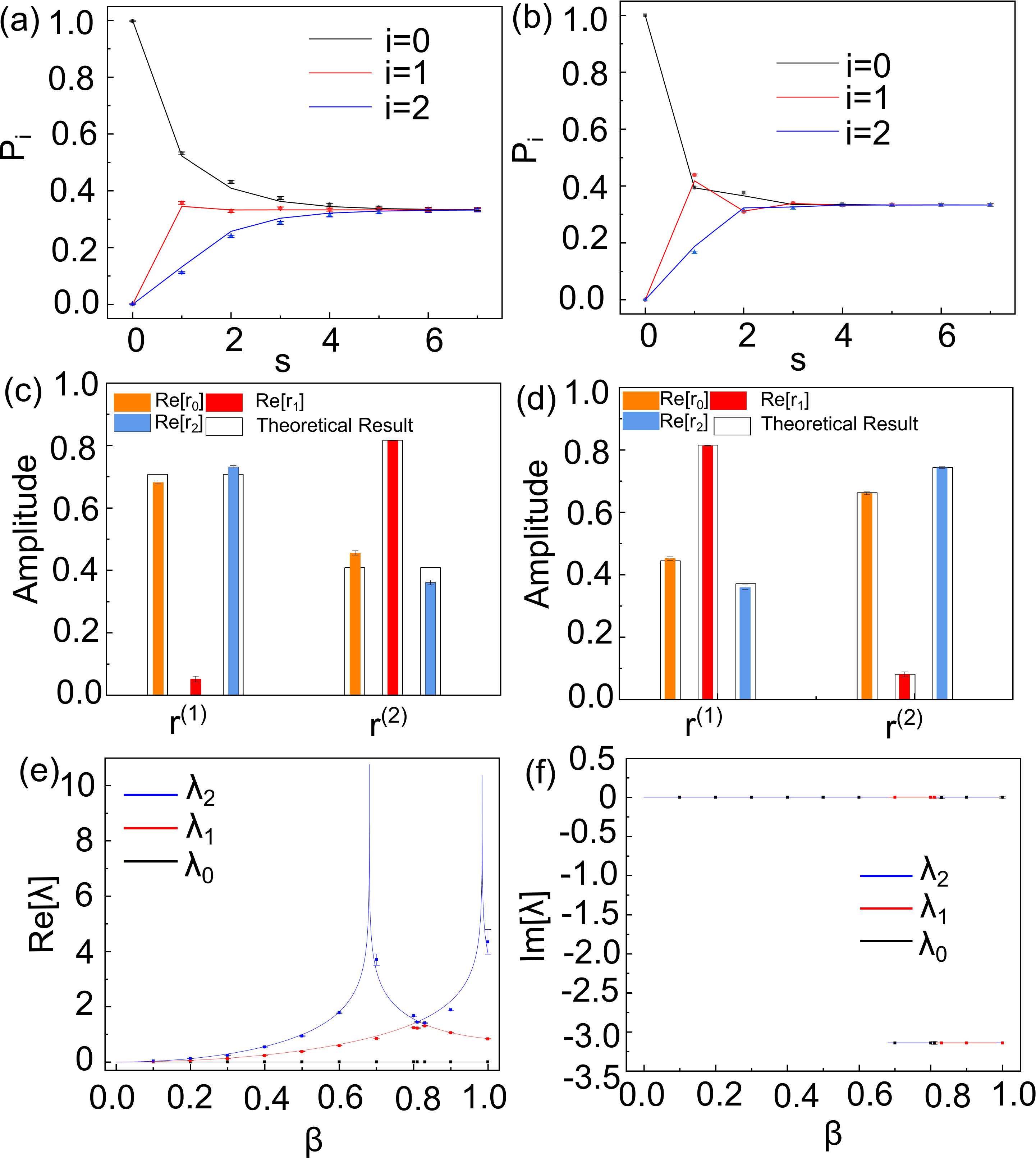}
	\caption{First-order dynamical phase transition in the fully dephased benchmark ($q=1$) under TRS ($\phi=0$).
		(a),(b) Measured step-resolved populations $P_i(s)$ ($i=0,1,2$) at $\beta=0.70$ ($<\beta_c$) and $\beta=0.83$ ($>\beta_c$), showing a qualitative change of the relaxation pattern.
		(c),(d) Corresponding components of the two non-stationary right eigenmodes \(r^{(1)}\) and \(r^{(2)}\) of the reconstructed transition matrix \(Q\), labeled as the slow and fast relaxation modes according to increasing \(\mathrm{Re}[\lambda_k]\)
		(solid bars: experiment with $1\sigma$ error bars; open rectangles: theory).
	(e) Real parts $\mathrm{Re}[\lambda_k]$ ($k=0,1,2$) of the Floquet exponents extracted from $Q(\beta)$, exhibiting a diabolic crossing $\mathrm{Re}[\lambda_1]=\mathrm{Re}[\lambda_2]$ at $\beta_c$.
		(f) Imaginary parts $\mathrm{Im}[\lambda_k]$, which remain pinned (0 or $\pi$ in the principal-branch convention) and do not develop a conjugate pair under TRS.}
	
	\label{fig2}
\end{figure}

\emph{First-order phase transition under TRS.}
We first consider the benchmark case \(q=1\) through its direct dynamical signatures. For \(\beta=0.70\), the population of node~1 reaches its stationary value already at \(s\simeq 2\), whereas the populations of nodes~0 and~2 relax on a longer timescale [Fig.~\ref{fig2}(a)]. This apparent node-dependent relaxation does not reflect different intrinsic decay rates of the nodes; rather, it arises from the different projections of the initial state onto the relaxation eigenmodes of the dynamics. To make this explicit, we express the deviation of the node population from the stationary state as
\[
\delta P_i(s)\equiv P_i^{(s)}-P_i^{\rm ss}
=\sum_{\ell=1}^{2} c_\ell E_\ell^{\,s} r_i^{(\ell)},
\]
 where \(P_i^{\rm ss}=1/3\), and \(E_\ell\), \(r^{(\ell)}\), and \(c_\ell\) denote the non-stationary eigenvalues, right eigenvectors, and overlap coefficients with the initial state, respectively. A node whose dynamics has only a weak projection onto the slow relaxation mode \(r^{(1)}\) therefore appears to equilibrate rapidly, whereas a larger projection gives rise to a pronounced long-time tail.

Reconstruction of the transition matrix \(Q\) fully confirms this interpretation (see the Supplemental Material~\cite{sm} for details of the reconstruction procedure and uncertainty analysis). At \(\beta=0.70\), \(r^{(1)}\) has an almost vanishing node-1 component, so the long-time tail is mainly visible on nodes~0 and~2.
At \(\beta=0.83\), the branch identified as \(r^{(1)}\) after the crossing acquires clear node-1 weight, and all three populations relax on comparable time scales [Fig.~\ref{fig2}(b),(d)].
The extracted Floquet exponents show a diabolic crossing \(\mathrm{Re}[\lambda_1]=\mathrm{Re}[\lambda_2]\) at \(\beta_c\), while \(\mathrm{Im}[\lambda_k]\) remains pinned to \(0\) or \(\pi\) under the principal-branch convention [Fig.~\ref{fig2}(e),(f)].
Across the crossing, the slow-relaxation branch switches abruptly [Fig.~\ref{fig2}(c),(d)], producing the discontinuous relaxation pattern that defines the FOPT.

To quantify this discontinuous switch with a single scalar observable, we define a relaxation order parameter from the slow relaxation mode,
\begin{equation}
	m(\beta)\equiv \big|r^{(\mathrm{slow})}_{1}(\beta)\big|^{2},\qquad
	\sum_{i=0}^{2}\big|r^{(\mathrm{slow})}_{i}(\beta)\big|^{2}=1,
	\label{eq:order_parameter_m}
\end{equation}
where \(r^{(\mathrm{slow})}\equiv r^{(1)}\), and \(r^{(\mathrm{slow})}_{i}\) denotes its component on node \(i\). Thus, \(m(\beta)\) measures the node~$1$ weight of the slow relaxation mode.

The critical point $\beta_c$ is identified by the diabolic crossing $\mathrm{Re}[\lambda_1]=\mathrm{Re}[\lambda_2]$, and we define the jump size as
$\Delta m \equiv m(\beta_c^{+})-m(\beta_c^{-})$. In practice, $m(\beta_c^{-})$ is averaged over $\beta=0.80,0.81$ and $m(\beta_c^{+})$ over $\beta=0.82,0.83$.
From the reconstructed $Q(\beta)$, we obtain $\beta_c=0.82\pm0.01$ and $\Delta m=0.662\pm0.008$.
The continuous $m(\beta)$ curve and the uncertainty analysis are given in Supplemental Material~\cite{sm}.

\begin{figure}[htp]
	\centering
	\includegraphics[width=\linewidth]{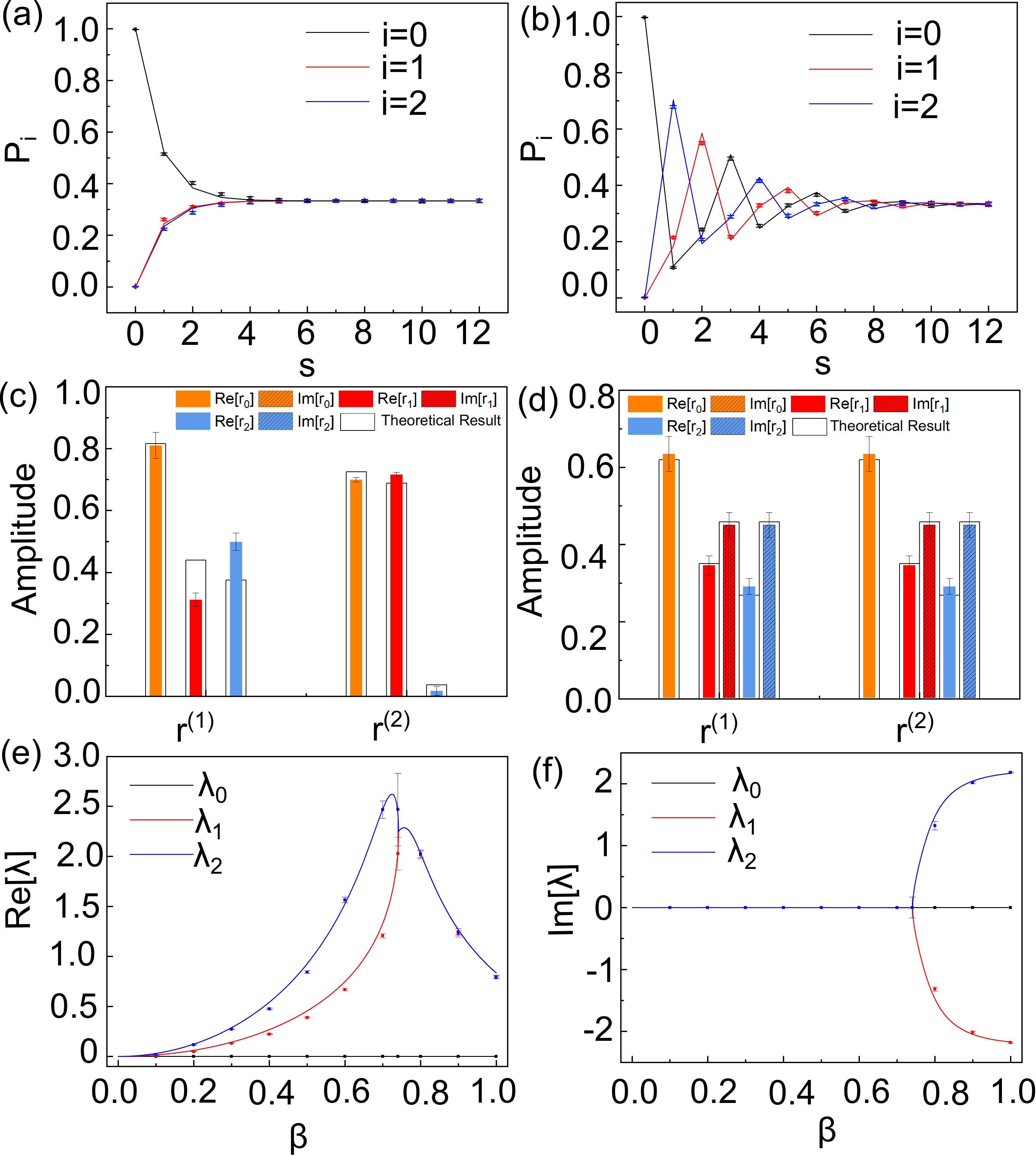}
	\caption{Second-order dynamical phase transition in the fully dephased benchmark ($q=1$) with broken TRS ($\phi=\pi/3$).
	 (a),(b) Measured populations \(P_i(s)\) at \(\beta=0.70<\beta_c\) and \(\beta=1.20>\beta_c\), showing monotonic and damped-oscillatory relaxation.
			(c),(d) Real and imaginary components of the two non-stationary right eigenmodes \(r^{(1)}\) and \(r^{(2)}\) of \(Q\) (solid bars: experiment with error bars; open rectangles: theory).
			(e) \(\mathrm{Re}[\lambda_k]\) versus \(\beta\), showing coalescence of the two non-stationary exponents at \(\beta_c\).
			(f) \(\mathrm{Im}[\lambda_k]\), showing the emergence of a complex-conjugate pair for \(\beta>\beta_c\).}
	\label{fig3}
\end{figure}

\begin{figure*}[t]
	\centering
	\includegraphics[width=1\linewidth]{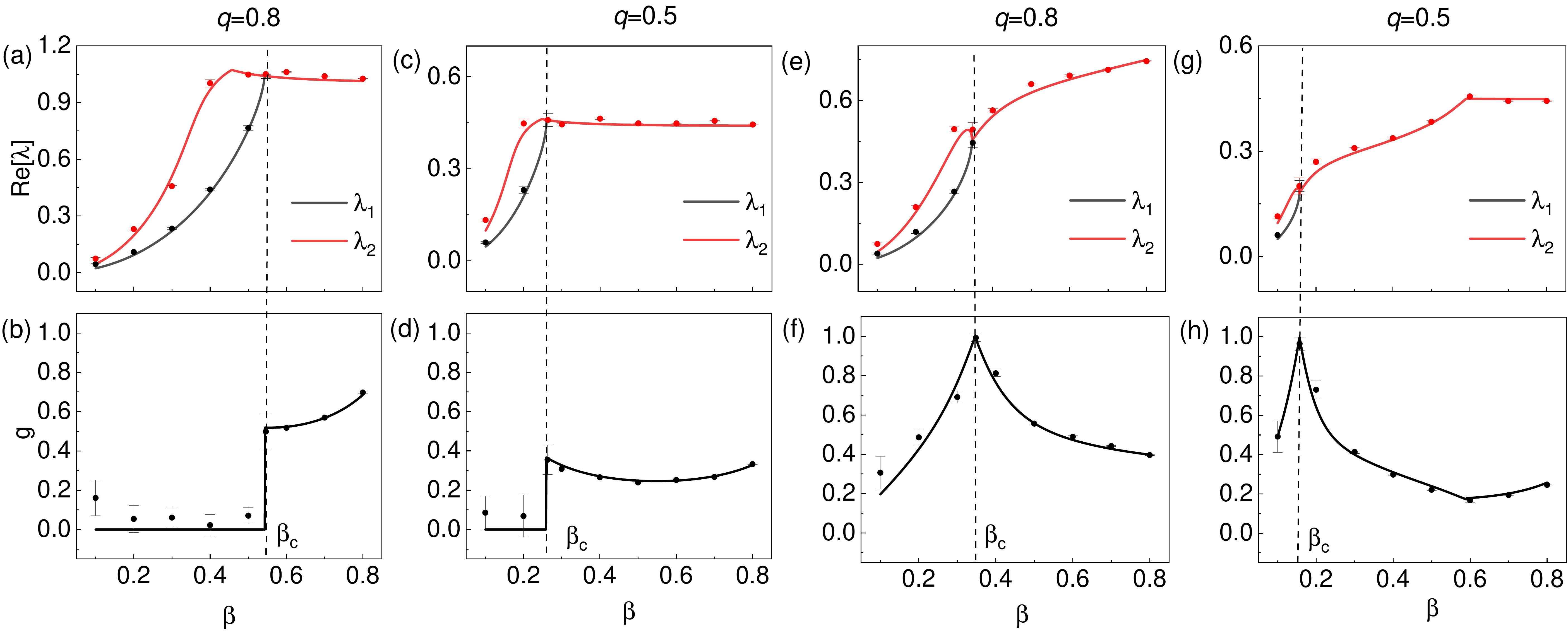}
	\caption{Quantum-regime validation in the OQW.
	Experimentally measured $\mathrm{Re}[\lambda_{1,2}]$ (top row) and eigenvector overlap $g$ (bottom row) versus $\beta$ at representative dephasing strengths.
		(a),(b) TRS case ($\phi=0$) at $q=0.8$: a real-eigenvalue crossing signals an FOPT while $g<1$ confirms distinct eigenmodes.
		(c),(d) Same as (a),(b) but for $q=0.5$.
		(e),(f) Broken-TRS case ($\phi=\pi/3$) at $q=0.8$: eigenvalues and eigenvectors coalesce at an EP, yielding an SOPT with $g\to 1$.
		(g),(h) Same as (e),(f) but for $q=0.5$.
		Vertical dashed lines indicate $\beta_c$.
		Solid curves are theoretical predictions; symbols are experimental data with photon-counting error bars.}
	\label{fig4}
\end{figure*}

\emph{Second-order phase transitions under broken TRS.}
We next turn to the TRS-broken setting, where detailed balance is violated and $Q$ is generally non-symmetric.
Experimentally, the relaxation changes from monotonic decay to damped oscillations:
	for \(\beta=0.70<\beta_c\) the populations relax monotonically [Fig.~\ref{fig3}(a)], whereas for \(\beta=1.20>\beta_c\) clear oscillations are resolved [Fig.~\ref{fig3}(b)]. Consistently, the reconstructed non-stationary modes are essentially real below \(\beta_c\), but form a complex-conjugate pair above it [Fig.~\ref{fig3}(c),(d)].
	The spectrum shows that \(\lambda_1\) and \(\lambda_2\) coalesce at \(\beta_c\simeq0.74\): their real parts merge, while their imaginary parts bifurcate with opposite signs [Fig.~\ref{fig3}(e),(f)].
	Together with eigenvector coalescence, this identifies an EP and establishes the SOPT mechanism. A flux-dependent scan further confirms this mechanism: the real diabolic crossing at \(\phi=0\) is converted into an EP at finite \(|\phi|\), beyond which the modes form a complex-conjugate pair (details are given in Supplemental Material~\cite{sm}).
	
The continuous evolution of the eigenmodes across the EP therefore yields a smooth change in the relaxation character,
distinct from the abrupt switching of the slow-relaxation branch in the FOPT. In particular, combining the complex-conjugate contributions yields real long-time deviations of the generic form
	$\delta P_i(s)=e^{-\kappa s}\big[A_i\cos(\omega s)+B_i\sin(\omega s)\big]$, with $A_i,B_i\in\mathbb{R}$, where we parameterize the conjugate pair as
	$\lambda_{\pm}=\kappa\pm i\omega$ with $\kappa>0$. To rule out technical oscillations, we perform a TRS control measurement at $\phi=0$ using the same $(\beta,q)$ and find no resolvable oscillatory component. Moreover, the fitted frequency $\omega_{\rm fit}$ agrees with the spectral prediction $\omega_{\rm spec}\equiv|\mathrm{Im}[\lambda_{\pm}]|$ extracted from the reconstructed spectrum (Supplemental Material~\cite{sm}, Sec.~\uppercase\expandafter{\romannumeral 6}).

To quantify eigenvector coalescence, we introduce the normalized overlap between the two non-stationary right eigenmodes,
\begin{equation}
	g(\beta)\equiv
	\frac{\left|\left(r^{(1)}\right)^{\dagger} r^{(2)}\right|}
	{\sqrt{\left[\left(r^{(1)}\right)^{\dagger} r^{(1)}\right]
			\left[\left(r^{(2)}\right)^{\dagger} r^{(2)}\right]}} ,
	\label{eq:g_def_main}
\end{equation}
where $r^{(1)}$ and $r^{(2)}$ are the right eigenvectors associated with $\lambda_{1}$ and $\lambda_{2}$
(i.e., $\lambda_{1,2}\equiv\lambda_{\pm}$ in the broken-TRS regime).
With this normalization, $0\le g\le 1$, and $g\to 1$ at an EP.
Near $\beta_c$ in the broken-TRS case, the two non-stationary modes exhibit a Puiseux expansion,
$\lambda_{1,2}(\beta)\simeq \lambda_{\rm EP}\pm a\sqrt{\beta-\beta_c}$,
so that the eigenvalue splitting $\Delta\lambda\equiv|\lambda_1-\lambda_2|$ vanishes as $\Delta\lambda\propto \sqrt{|\beta-\beta_c|}$.
For convenience we also define the eigenmode splitting $d_r\equiv \sqrt{1-g^2}$, which follows the same square-root scaling. Details of the Puiseux expansion and the square-root scaling analysis for $\Delta\lambda$ and $d_r$ are provided in the End Matter and Supplemental Material~\cite{sm}.

\emph{Quantum-regime validation.}
In the full-quantum regime \(0<q<1\), coherences are only partially damped and the relaxation dynamics is governed by the spectrum of the one-step superoperator \(S\equiv e^{-\mathcal{L}}\).
We therefore extend our measurements into the \(0<q<1\) regime and reconstruct \(S\) to track the two leading non-stationary exponents \(\lambda_{1,2}\) and their eigenvector overlap \(g\) as functions of \(\beta\) (see Supplemental Material~\cite{sm}). Figure~\ref{fig4} shows representative results for \(q=0.8\) and \(q=0.5\). In each case, a well-defined critical point \(\beta_c\) is identified by the crossing or coalescence of \(\mathrm{Re}[\lambda_1]\) and \(\mathrm{Re}[\lambda_2]\) [Fig.~\ref{fig4}(a),(c),(e),(g)], accompanied by a pronounced variation of the overlap parameter \(g\) [Fig.~\ref{fig4}(b),(d),(f),(h)], closely paralleling the classical (\(q=1\)) behavior discussed above.
At \(\beta_c\), the TRS-preserving case [Fig.~\ref{fig4}(b),(d)] exhibits \(g<1\), consistent with a diabolic crossing, whereas in the TRS-broken case [Fig.~\ref{fig4}(f),(h)] \(g\) approaches unity, indicating eigenvector coalescence at an EP. A comparison between the \(q=0.8\) and \(q=0.5\) data sets further reveals a systematic shift of \(\beta_c\) with dephasing strength. As coherence increases (i.e. as \(q\) decreases), the critical point moves monotonically toward smaller values of \(\beta\). For weak dephasing (\(q=0.1\)), both the spectral and eigenvector signatures become strongly smeared out (End Matter, Fig.~\ref{fig:EM_qscan}), indicating that the transition is progressively suppressed in the near-unitary limit.

\emph{Discussion}.
In summary, we experimentally demonstrate controllable first- and second-order dynamical phase transitions in the relaxation dynamics of an OQW within a single photonic platform, providing a first experimental realization of dynamical criticality beyond the closed-system paradigm based on quench-induced nonanalyticities.
While phase transitions are typically associated with the thermodynamic limit in dissipative systems \cite{minganti2018spectral} or quenches in closed many-body systems \cite{Heyl_Review_2018}, our results show that sharp dynamical critical behavior can arise even in a finite open system and is governed by the spectral structure of the Floquet--Liouvillian superoperator.
In particular, the direct observation of an EP in the broken-TRS regime provides experimental access to eigenvalue and eigenmode coalescence in dissipative dynamics, revealing complex spectral structures forbidden under TRS.
These observations establish a direct connection between quantum-induced non-Hermiticity and the asymptotic relaxation of stochastic processes, and show that even near the classical limit ($q\approx1$) the universality class of the relaxation dynamics is controlled by the synthetic gauge flux $\phi$.

Beyond fundamental physics, the ability to switch the order of the dynamical transition using a tunable gauge phase provides a practical handle for spectrum-guided relaxation engineering; operation near the critical point $\beta_c$ also enhances parametric sensitivity associated with the EP singularity.
Our platform thus opens a route to programmable dissipative dynamics and controlled exploration of non-Hermitian criticality in quantum simulators.
Extending the scheme to larger photonic networks could enable studies of non-Hermitian spectral topology in open systems, including bulk--boundary correspondence and dissipative topological phases, and motivate searches for boundary-sensitive effects such as the Liouvillian non-Hermitian skin effect~\cite{Song_PRL_2019,Okuma_PRL_2020,Mao_PRB_2024} and non-Markovian EPs~\cite{Longhi2026}.
Further directions include exploring the continuous-time limit and alternative decoherence mechanisms, as well as exploiting tunable EPs to tailor or accelerate relaxation, enabling tests of non-Hermitian speedup phenomena such as Mpemba-like effects~\cite{uff2,uff3,Ares_NRP_2025,Joshi_PRL_2024} and fast state preparation in scalable photonic simulators.

\clearpage
\appendix
\onecolumngrid
\subsection*{End Matter}
\twocolumngrid

\begin{figure*}[htbp]
	\centering
	\includegraphics[width=0.75\linewidth]{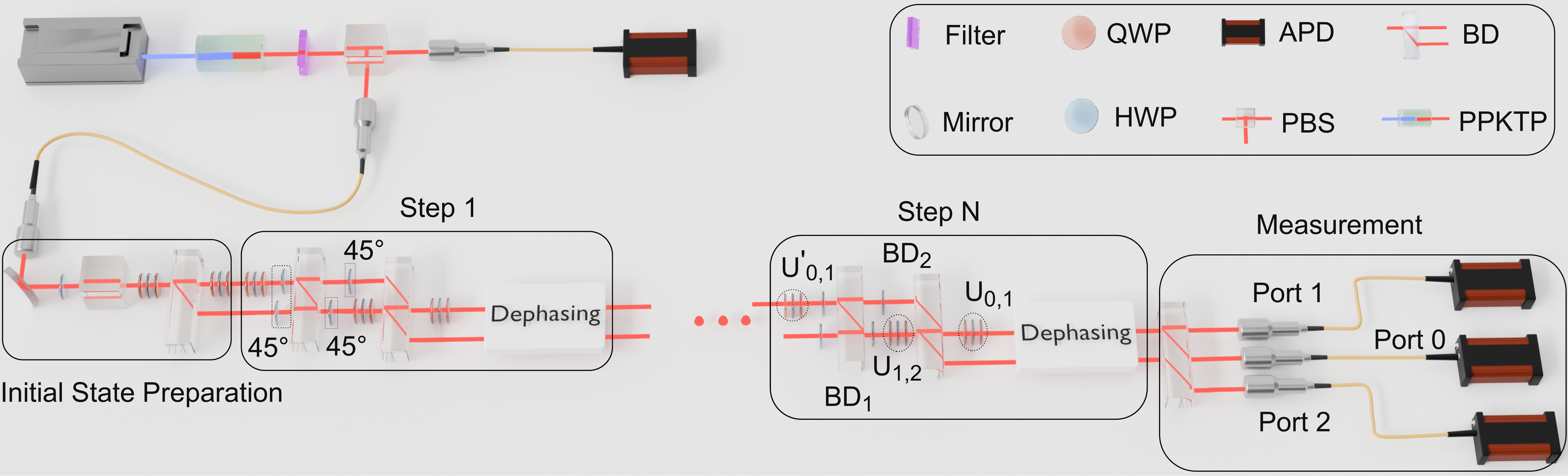}
	\caption{Experimental schematic.
		Each discrete-time step of the OQW consists of a coherent unitary operation $\mathcal{U}$ followed by a dephasing channel.
		Beam displacers map the three path modes to spatially separated ports, which are detected by APDs to yield node-resolved counts.
		Thin outlined boxes indicate $45^{\circ}$ HWPs used for mode readdressing between the two BDs.
		Dashed ellipses mark QHQ sets implementing the two-mode unitaries $U_{i,j}$ in the decomposition of $\mathcal{U}$.
		The optical synthesis of the full $3\times 3$ step unitary $\mathcal{U}$ is shown in the solid box.}
	\label{fig:EM_setup}
\end{figure*}

\begin{figure*}[htbp]
	\centering
	\includegraphics[width=0.6\linewidth]{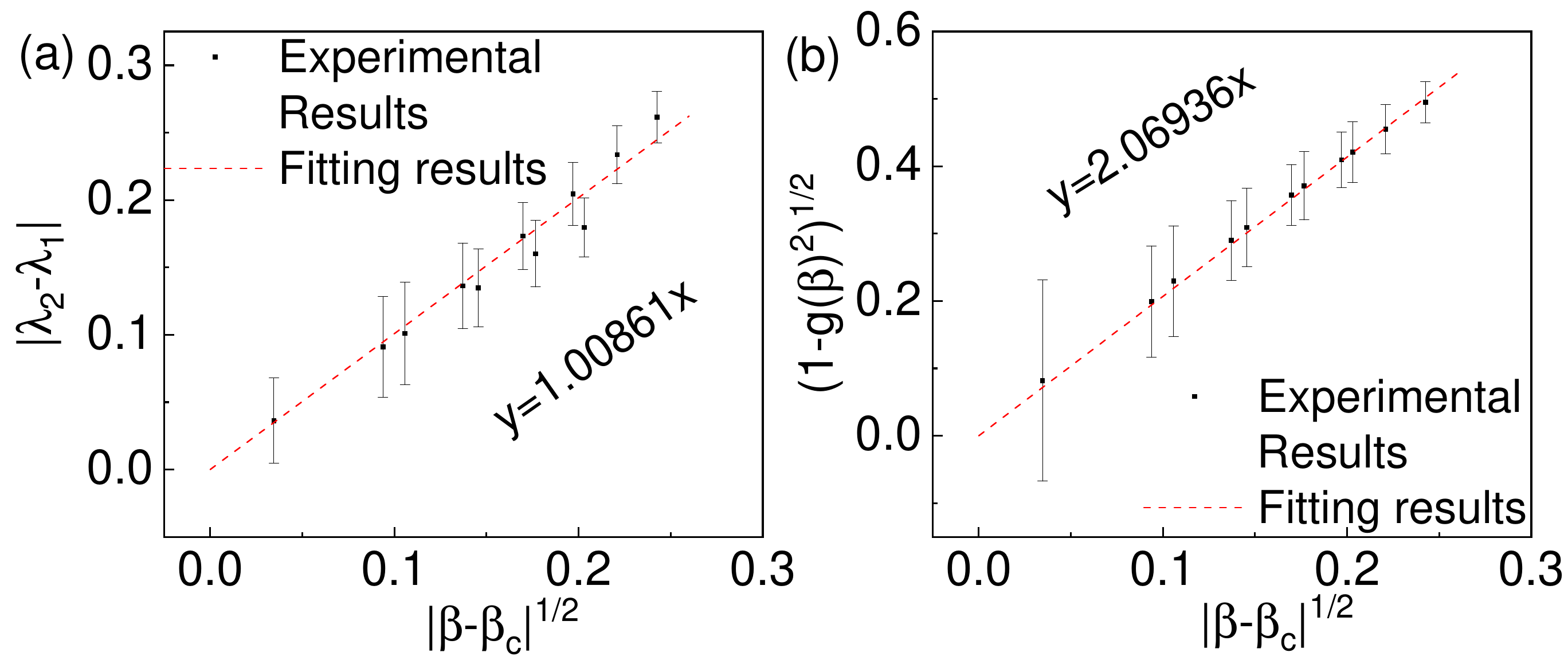}
	\caption{Square-root scaling near the EP.
		(a) Eigenvalue splitting $\Delta\lambda=|\lambda_+-\lambda_-|$ versus
		$x=\sqrt{|\beta-\beta_c|}$.
		(b) Eigenmode splitting $d_r=\sqrt{1-g^2}$ versus $x$.
		Solid lines are weighted linear fits. Error bars denote one standard deviation propagated from photon-counting statistics.}
	\label{fig:EM_scaling}
\end{figure*}

\begin{figure*}[htbp]
	\centering
	\includegraphics[width=1\linewidth]{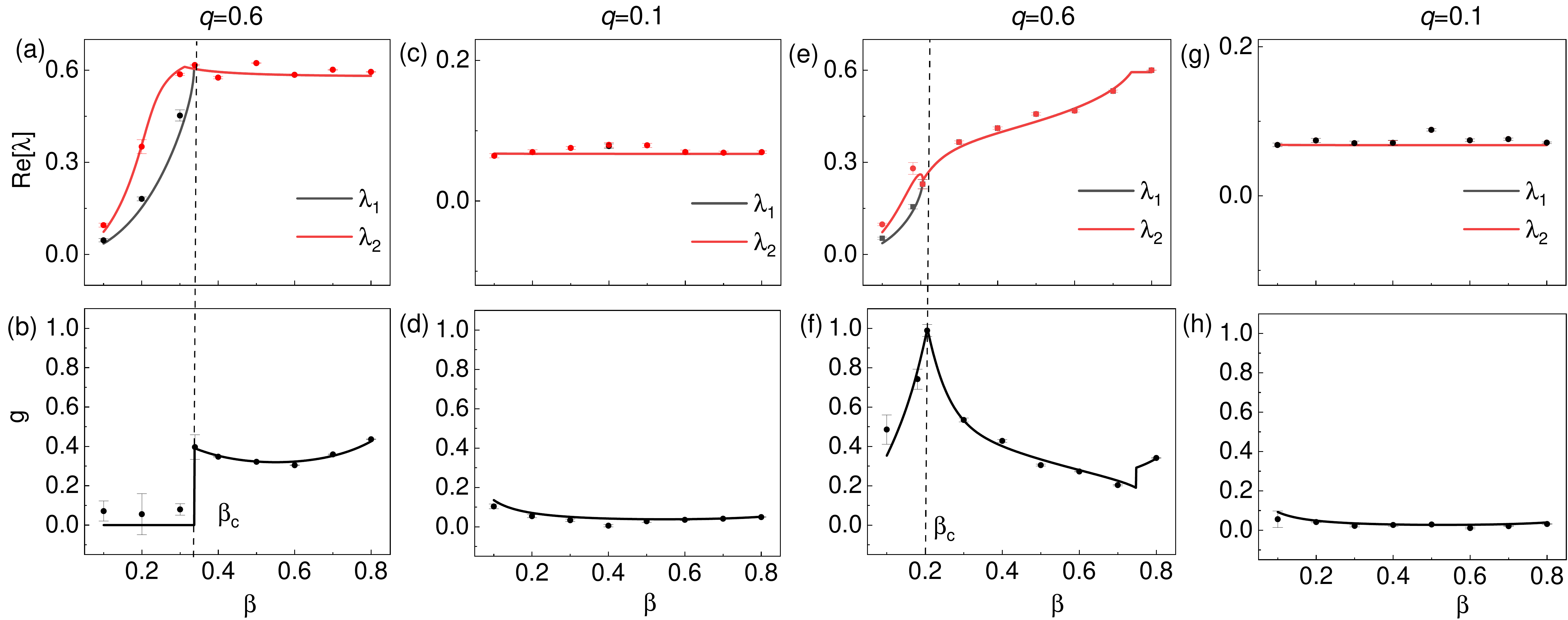}
	\caption{Complementary scans deeper into the quantum regime.
		Same observables as Fig.~\ref{fig4}, shown for $q=0.6$ and $0.1$.
		(a),(b) TRS case ($\phi=0$) at $q=0.6$: the real-eigenvalue crossing persists (FOPT) and $g$ remains below unity.
		(c),(d) TRS case at $q=0.1$: no resolvable spectral crossing is observed and $g$ shows no critical feature, indicating suppressed criticality.
		(e),(f) Broken-TRS case ($\phi=\pi/3$) at $q=0.6$: the EP-driven SOPT remains visible through simultaneous eigenvalue and eigenvector coalescence ($g\to 1$).
		(g),(h) Broken-TRS case at $q=0.1$: the EP signature and the corresponding peak in $g$ are washed out.
		Vertical dashed lines indicate $\beta_c$ where a critical point is identifiable.}
	\label{fig:EM_qscan}
\end{figure*}

{\it Appendix A: Experimental setup.---} Figure~\ref{fig:EM_setup} shows the full single-photon interferometric network used to implement one discrete step of the OQW.
Heralded single photons are generated by type-II spontaneous parametric down-conversion in a 20-mm-long PPKTP crystal pumped by a 405-nm continuous-wave diode laser. One photon serves as a trigger, while the other is injected into the OQW.
After the polarization beam splitter (PBS) prepares the reference state $\ket{UH}$, a BD creates the upper ($U$) and lower ($D$) spatial modes.
An arbitrary qutrit input state
\[
a_0\ket{UH}+a_1 e^{i\varphi_1}\ket{UV}+a_2 e^{i\varphi_2}\ket{DH}
\]
is prepared by half-wave plates $H_0$ and $H_1$ for the amplitudes and by quarter-wave plate--half-wave plate--quarter-wave plate (QHQ) sets for the phases $\varphi_1$ and $\varphi_2$; the explicit angle conventions are given in the Supplemental Material~\cite{sm}.

To match the physical layout of the interferometer, we factorize the three-mode step unitary as
	\begin{equation}
		\mathcal{U}=U_{0,1}\,U_{1,2}\,U'_{0,1},
		\label{uu}
	\end{equation}
	where each $U_{i,j}$ denotes an $\mathrm{SU}(2)$ operation acting only on the two-dimensional subspace spanned by modes $|i\rangle$ and $|j\rangle$, while leaving the third mode unchanged.
	Each such two-mode unitary is implemented by a QHQ set.
	In the optical network, the three QHQ sets before, between, and after the two BDs implement $U'_{0,1}$, $U_{1,2}$, and $U_{0,1}$, respectively.
	Two $45^\circ$ half-wave plates (HWPs) between the BDs readdress the mode labels so that the coupled subspace is switched from $(0,1)$ to $(1,2)$ and back to $(0,1)$.
	The corresponding wave-plate angles are obtained numerically from the target two-mode blocks, as detailed in the Supplemental Material~\cite{sm}.
	
	Dephasing is implemented by two complementary modules optimized for different parameter regimes.
	For the fully dephased benchmark, fine control of the dephasing strength is unnecessary: it is sufficient to suppress interference among the three modes after each step.
	We therefore use a compact polarization-delay interferometer built from two PBSs, which is experimentally simpler, more space-efficient, and reliably yields $q\simeq 1$ by washing out the $H$--$V$ coherence within each step.
	For $0<q<1$, by contrast, controlled partial dephasing is required.
	We therefore use a two-BD interferometer with a tunable relative tilt that introduces adjustable mode mismatch, following the BD-tilt strategy of Ref.~\cite{broome2010discrete}.
	In both cases, the effective dephasing strength $q$ is calibrated from the single-photon interference visibility rather than inferred directly from the mechanical setting alone.
	After each step, a final BD maps the three modes to spatially separated output ports, and APDs record the node-resolved photon counts.

{\it Appendix B: Exceptional-point square-root scaling.---}
A defining signature of a second-order EP is the square-root (Puiseux) criticality of both
eigenvalues and eigenmodes as the control parameter approaches the EP.
Using the experimentally reconstructed spectrum, we evaluate the eigenvalue splitting
$\Delta\lambda(\beta)\equiv|\lambda_{2}(\beta)-\lambda_{1}(\beta)|$
and the eigenmode splitting
$d_r(\beta)\equiv \sqrt{1-g(\beta)^2}$,
where $g(\beta)$ is the right-eigenvector overlap defined in the main text.
Near the EP at $\beta_c$, both quantities obey
\begin{equation}
	\Delta\lambda(\beta)\sim A_\lambda \sqrt{|\beta-\beta_c|},\qquad
	d_r(\beta)\sim A_r \sqrt{|\beta-\beta_c|}.
	\label{eq:EP_sqrt_scaling}
\end{equation}

As shown in Fig.~\ref{fig:EM_scaling}, the data are linear when plotted versus
$x\equiv\sqrt{|\beta-\beta_c|}$.
Weighted fits constrained through the origin give
	\(A_\lambda=1.00\pm0.02\) and \(A_r=2.069\pm0.009\).
Allowing a free intercept leaves $\Delta\lambda$ consistent with zero offset, while $d_r$ exhibits a small residual offset,
consistent with finite experimental resolution when $g\to 1$.
Additional fit models, goodness-of-fit, and uncertainty-propagation details are provided in the Supplemental Material~\cite{sm}.

{\it Appendix C: Phase transitions in the full quantum regime.---}
Having established the transition mechanism in the strong-dephasing benchmark ($q=1$), we probe the hybrid quantum--classical regime $0<q<1$, where coherent evolution and dissipation compete.
We tune $q$ continuously using the BD-tilt scheme and calibrate it from single-photon interference visibility measurements (see Supplemental Material~\cite{sm}).
For each $(\beta,\phi,q)$, we reconstruct the one-step Floquet--Liouvillian superoperator and extract the two leading non-stationary exponents \(\lambda_{1,2}\), ordered by increasing \(\mathrm{Re}[\lambda_k]\), together with the eigenvector-overlap parameter \(g\) (defined in the main text).
Representative scans at $q=0.8$ and $0.5$ are shown in the main text (Fig.~\ref{fig4}).
Below we present complementary measurements at $q=0.6$ and $q=0.1$ (Fig.~\ref{fig:EM_qscan}). These results clearly indicate a monotonic drift of $\beta_c$ toward smaller $\beta$ as coherence increases and the disappearance of both FOPT/SOPT signatures in the near-unitary limit ($q=0.1$, Fig.~\ref{fig:EM_qscan}(c,d) and (g,h)).


\begin{thebibliography}{100}
\bibitem{Heyl_Review_2018}
M. Heyl, Dynamical quantum phase transitions: a review, Rep. Prog. Phys. \textbf{81}, 054001 (2018).

\bibitem{Heyl_PRL_2013}
M. Heyl, A. Polkovnikov, and S. Kehrein, Dynamical quantum phase transitions in the transverse-field Ising model, Phys. Rev. Lett. \textbf{110}, 135704 (2013).

\bibitem{Wang_PRL_2019}
K. Wang, X. Qiu, L. Xiao, X. Zhan, Z. Bian, W. Yi, and P. Xue, Simulating dynamic quantum phase transitions in photonic quantum walks, Phys. Rev. Lett. \textbf{122}, 020501 (2019).

\bibitem{Marino_2022}
J. Marino, M. Eckstein, M. S. Foster, and A. M. Rey, Dynamical phase transitions in the collisionless pre-thermal states of isolated quantum systems: Theory and experiments, Rep. Prog. Phys. \textbf{85}, 116001 (2022).



\bibitem{Fitzpatrick2017}
M.~Fitzpatrick, N.~M.~Sundaresan, A.~C.~Y.~Li, J.~Koch, and A.~A.~Houck,
Observation of a dissipative phase transition in a one-dimensional circuit QED lattice,
Phys. Rev. X \textbf{7}, 011016 (2017).

\bibitem{Rodriguez2017}
S.~R.~K.~Rodriguez, W.~Casteels, F.~Storme, N.~Carlon~Zambon,
I.~Sagnes, L.~Le~Gratiet, E.~Galopin, A.~Lema\^{i}tre, A.~Amo,
C.~Ciuti, and J.~Bloch,
Probing a dissipative phase transition via dynamical optical hysteresis,
Phys. Rev. Lett. \textbf{118}, 247402 (2017).

\bibitem{Fink2018}
T.~Fink, A.~Schade, S.~H\"ofling, C.~Schneider, and A.~Imamoglu,
Signatures of a dissipative phase transition in photon correlation measurements,
Nat.\ Phys.\ \textbf{14}, 365 (2018).


\bibitem{minganti2018spectral}
F. Minganti, A. Biella, N. Bartolo, and C. Ciuti, Spectral theory of Liouvillians for dissipative phase transitions, Phys. Rev. A \textbf{98}, 042118 (2018).




\bibitem{Gao_PRL_2024}
H. Gao, K. Wang, L. Xiao, M. Nakagawa, N. Matsumoto, D. Qu, H. Lin, M. Ueda, and P. Xue, Experimental observation of the Yang-Lee quantum criticality in open quantum systems, Phys. Rev. Lett. \textbf{132}, 176601 (2024).

\bibitem{Zunkovic_PRL_2018}
B. \v{Z}unkovi\v{c}, M. Heyl, M. Knap, and A. Silva, Dynamical quantum phase transitions in spin chains with long-range interactions: Merging different concepts of nonequilibrium criticality, Phys. Rev. Lett. \textbf{120}, 130601 (2018).



\bibitem{Teza_PRL_2023}
G. Teza, R. Yaacoby, and O. Raz, Eigenvalue crossing as a phase transition in relaxation dynamics, Phys. Rev. Lett. \textbf{130}, 207103 (2023).

\bibitem{mori2023symmetrized}
T. Mori and T. Shirai, Symmetrized Liouvillian gap in Markovian open quantum systems, Phys. Rev. Lett. \textbf{130}, 230404 (2023).


\bibitem{casteels2017critical}
W. Casteels, R. Fazio, and C. Ciuti, Critical scaling of the Liouvillian gap for a nonlinear driven-dissipative resonator, Phys. Rev. A \textbf{95}, 012128 (2017).

\bibitem{Blythe_JPCS_2006}
R. A. Blythe, An introduction to phase transitions in stochastic dynamical systems, J. Phys.: Conf. Ser. \textbf{40}, 1 (2006).

\bibitem{xiao2024non}
L. Xiao, Y. Chu, Q. Lin, H. Lin, W. Yi, J. Cai, and P. Xue, Non-Hermitian sensing in the absence of exceptional points, Phys. Rev. Lett. \textbf{133}, 180801 (2024).

\bibitem{shi2025enhanced}
T. Shi, V. Smirnov, K. Shi, and W. Zhang, Enhanced response at exceptional points in multiqubit systems for sensing, Phys. Rev. A \textbf{111}, 032203 (2025).

\bibitem{arkhipov2021generating}
I. I. Arkhipov, F. Minganti, A. Miranowicz, and F. Nori, Generating high-order quantum exceptional points in synthetic dimensions, Phys. Rev. A \textbf{104}, 012205 (2021).

\bibitem{xue2020non}
H. Xue, Q. Wang, B. Zhang, and Y. D. Chong, Non-Hermitian Dirac cones, Phys. Rev. Lett. \textbf{124}, 236403 (2020).

\bibitem{lamb1998time}
J. S. W. Lamb and J. A. G. Roberts, Time-reversal symmetry in dynamical systems: A survey, Physica D \textbf{112}, 1 (1998).

\bibitem{choi2023noninvertible}
Y. Choi, H. T. Lam, and S.-H. Shao, Noninvertible time-reversal symmetry, Phys. Rev. Lett. \textbf{130}, 131602 (2023).

\bibitem{sato2012time}
M. Sato, K. Hasebe, K. Esaki, and M. Kohmoto, Time-reversal symmetry in non-Hermitian systems, Prog. Theor. Phys. \textbf{127}, 937 (2012).

\bibitem{parzygnat2023time}
A. J. Parzygnat and J. Fullwood, From time-reversal symmetry to quantum Bayes' rules, PRX Quantum \textbf{4}, 020334 (2023).

\bibitem{uff1}
G. Teza, R. Yaacoby, and O. Raz, Relaxation shortcuts through boundary coupling,
Phys. Rev. Lett. \textbf{131}, 017101 (2023).

\bibitem{uff2}
M. Moroder, O. Culhane, K. Zawadzki, and J. Goold, Thermodynamics of the quantum Mpemba effect, Phys. Rev. Lett. \textbf{133}, 140404 (2024).

\bibitem{uff3}
G. Teza, J. Bechhoefer, A. Lasanta, O. Raz, and M. Vucelja, Speedups in nonequilibrium thermal relaxation: Mpemba and related effects, Phys. Rep. \textbf{1164}, 1--97 (2026).

\bibitem{Longhi_AQT_2025}
S. Longhi, Dynamical phase transitions in open quantum walks, Adv. Quantum Technol. \textbf{8}, e00539 (2025).

\bibitem{uffa4}
I. Ohnishi, Exceptional points in dephased topological quantum walks as Lyapunov stable non-equilibrium attractors, SSRN \textbf{5785442} (2025).


\bibitem{BH2024}
B. Huo, Q. Lin, L. Xiao, X. Zhan, D. Qu, and P. Xue, Non-Bloch parity-time-symmetric phase transition in quantum walks, Phys. Rev. A \textbf{110}, 052410 (2024).



\bibitem{Kwiat_Science_2000}
P. G. Kwiat, A. J. Berglund, J. B. Altepeter, and A. G. White,
Experimental verification of decoherence-free subspaces,
Science \textbf{290}, 498--501 (2000).

\bibitem{Kendon_PRA_2003}
V. Kendon and B. Tregenna,
Decoherence can be useful in quantum walks,
Phys. Rev. A \textbf{67}, 042315 (2003).

\bibitem{Brun_PRA_2003}
T. A. Brun, H. A. Carteret, and A. Ambainis,
Quantum random walks with decoherent coins,
Phys. Rev. A \textbf{67}, 032304 (2003).

\bibitem{Kendon_LNP_2004}
V. Kendon and B. Tregenna,
Decoherence in discrete quantum walks,
in \textit{Decoherence and entropy in complex systems},
edited by H.-T. Elze,
Lecture Notes in Physics Vol. \textbf{633}
(Springer, Berlin, Heidelberg, 2004), pp. 253--267.

\bibitem{Kendon_MSCS_2007}
V. Kendon,
Decoherence in quantum walks---a review,
Math. Struct. Comput. Sci. \textbf{17}, 1169--1220 (2007).

\bibitem{Annabestani_PRA_2010}
M. Annabestani, S. J. Akhtarshenas, and M. R. Abolhassani,
Decoherence in a one-dimensional quantum walk,
Phys. Rev. A \textbf{81}, 032321 (2010).


\bibitem{broome2010discrete}
M. A. Broome, A. Fedrizzi, B. P. Lanyon, I. Kassal, A. Aspuru-Guzik, and A. G. White, Discrete single-photon quantum walks with tunable decoherence, Phys. Rev. Lett. \textbf{104}, 153602 (2010).

\bibitem{Schreiber_PRL_2011}
A. Schreiber, K. N. Cassemiro, V. Poto\v{c}ek, A. G\'{a}bris, I. Jex, and C. Silberhorn, Decoherence and disorder in quantum walks: from ballistic spread to localization, Phys. Rev. Lett. \textbf{106}, 180403 (2011).

\bibitem{Chalabi_PRL_2019}
H. Chalabi, S. Barik, S. Mittal, T. E. Murphy, M. Hafezi, and E. Waks, Synthetic gauge field for two-dimensional time-multiplexed quantum random walks, Phys. Rev. Lett. \textbf{123}, 150503 (2019).


\bibitem{VenegasAndraca_QIP_2012}
S. E. Venegas-Andraca,
Quantum walks: a comprehensive review,
Quantum Inf. Process. \textbf{11}, 1015--1106 (2012).

\bibitem{Zhang_CPB_2013}
R. Zhang, H. Qin, B. Tang, and P. Xue,
Disorder and decoherence in coined quantum walks,
Chin. Phys. B \textbf{22}, 110312 (2013).

\bibitem{Alberti_NJP_2014}
A. Alberti, W. Alt, R. Werner, and D. Meschede,
Decoherence models for discrete-time quantum walks and their application to neutral atom experiments,
New J. Phys. \textbf{16}, 123052 (2014).

\bibitem{Caruso_NC_2016}
F. Caruso, A. Crespi, A. G. Ciriolo, F. Sciarrino, and R. Osellame,
Fast escape of a quantum walker from an integrated photonic maze,
Nat. Commun. \textbf{7}, 11682 (2016).
%\bibitem{Chatterjee_PRA_2024}
%A. K. Chatterjee, S. Takada, and H. Hayakawa, Multiple quantum Mpemba effect: Exceptional points and oscillations, Phys. Rev. A \textbf{110}, 022213 (2024).

\bibitem{Chuang_Nielsen_1997}
I. L. Chuang and M. A. Nielsen, Prescription for experimental determination of the dynamics of a quantum black box, J. Mod. Opt. \textbf{44}, 2455 (1997).

\bibitem{Poyatos_PRL_1997}
J. F. Poyatos, J. I. Cirac, and P. Zoller, Complete characterization of a quantum process: the two-bit quantum gate, Phys. Rev. Lett. \textbf{78}, 390 (1997).

\bibitem{OBrien_PRL_2004}
J. L. O'Brien, G. J. Pryde, A. Gilchrist, D. F. V. James, N. K. Langford, T. C. Ralph, and A. G. White, Quantum process tomography of a controlled-not gate, Phys. Rev. Lett. \textbf{93}, 080502 (2004).

\bibitem{Samach_PRApplied_2022}
G. O. Samach, A. Greene, J. Borregaard, M. Christandl, J. Barreto, D. K. Kim, C. M. McNally, A. Melville, B. M. Niedzielski, Y. Sung, D. Rosenberg, M. E. Schwartz, J. L. Yoder, T. P. Orlando, J. I.-J. Wang, S. Gustavsson, M. Kj{\ae}rgaard, and W. D. Oliver, Lindblad tomography of a superconducting quantum processor, Phys. Rev. Applied \textbf{18}, 064056 (2022).

\bibitem{Zhang_PRL_2019_EPNoise}
M. Zhang, W. Sweeney, C. W. Hsu, L. Yang, A. D. Stone, and L. Jiang, Quantum noise theory of exceptional point amplifying sensors, Phys. Rev. Lett. \textbf{123}, 180501 (2019).

\bibitem{Wiersig_PRA_2020}
J. Wiersig, Robustness of exceptional-point-based sensors against parametric noise: The role of Hamiltonian and Liouvillian degeneracies, Phys. Rev. A \textbf{101}, 053846 (2020).

\bibitem{Bergholtz_RMP_2021}
E.~J.~Bergholtz, J.~C.~Budich, and F.~K.~Kunst, Exceptional topology of non-Hermitian systems, Rev.~Mod.~Phys. \textbf{93}, 015005 (2021).

\bibitem{Okuma_ARCMP_2023}
N.~Okuma and M.~Sato, Non-Hermitian topological phenomena: A review, Annu.~Rev.~Condens.~Matter Phys. \textbf{14}, 83--107 (2023).

\bibitem{xiao2025non}
L. Xiao, K. Wang, D. Qu, H. Gao, Q. Lin, Z. Bian, X. Zhan, and P. Xue, Non-Hermitian physics in photonic systems, Photonics Insights \textbf{4}, R09 (2025).

\bibitem{DFM22}
K. Ding, C. Fang, and G. Ma, Non-Hermitian topology and exceptional-point geometries, Nat. Rev. Phys. \textbf{4}, 745--760 (2022).

\bibitem{Xue_PRL_2026} P. Xue, Essay: Topological phases and exceptional points in non-Hermitian systems, Phys. Rev. Lett. \textbf{136}, 170001 (2026).

\bibitem{Ares_NRP_2025}
F. Ares, P. Calabrese, and S. Murciano, The quantum Mpemba effects, Nat. Rev. Phys. \textbf{7}, 451--460 (2025).

\bibitem{Joshi_PRL_2024}
L. K. Joshi, J. Franke, A. Rath, F. Ares, S. Murciano, F. Kranzl, R. Blatt, P. Zoller, B. Vermersch, and P. Calabrese, Observing the quantum Mpemba effect in quantum simulations, Phys. Rev. Lett. \textbf{133}, 010402 (2024).

\bibitem{doublyS}
R. B. Bapat and T. E. S. Raghavan,
Doubly stochastic matrices, in \textit{Nonnegative Matrices and Applications},
\textit{Encyclopedia of Mathematics and its Applications}, pp.~59--114
(Cambridge University Press, Cambridge, UK, 1997).

\bibitem{sm}
See Supplemental Material.



%\bibitem{Zhang_NatCommun_2025}
%J. Zhang, G. Xia, C.-W. Wu, T. Chen, Q. Zhang, Y. Xie, W.-B. Su, W. Wu, C.-W. Qiu, P.-X. Chen, W. Li, H. Jing, Y.-L. Zhou, Observation of quantum strong Mpemba effect, Nat. Commun. \textbf{16}, 301 (2025).

%\bibitem{Berry_ProcRSocA_1984}
%M. V. Berry, Diabolical points in the spectra of triangles, Proc. R. Soc. Lond. A \textbf{392}, 45 (1984).

%\bibitem{Wiersig_PRR_2022}
%J. Wiersig, Distance between exceptional points and diabolic points and its implication for the response strength of non-Hermitian systems, Phys. Rev. Research \textbf{4}, 033179 (2022).

\bibitem{Song_PRL_2019}
F.~Song, S.~Yao, and Z.~Wang,
Non-Hermitian skin effect and chiral damping in open quantum systems,
Phys. Rev. Lett.~\textbf{123}, 170401 (2019).

\bibitem{Okuma_PRL_2020}
N.~Okuma, K.~Kawabata, K.~Shiozaki, and M.~Sato,
Topological origin of non-Hermitian skin effects,
Phys. Rev. Lett.~\textbf{124}, 086801 (2020).

\bibitem{Mao_PRB_2024}
L.~Mao, X.~Yang, M.~J.~Tao, H.~Hu, and L.~Pan,
Liouvillian skin effect in a one-dimensional open many-body system,
Phys.~Rev.~B~\textbf{110}, 045440 (2024).

\bibitem{Longhi2026}
S.~Longhi, Boundary-driven exceptional points in photonic waveguide lattices, Opt.~Lett. \textbf{51}, 1740--1743 (2026).


%\bibitem{Moroder_PRL_2024}
%M. Moroder, O. Culhane, K. Zawadzki, and J. Goold, Thermodynamics of the quantum Mpemba effect, Phys. Rev. Lett. \textbf{133}, 140404 (2024).



%————————————————————————————————SM ref
	
\end{thebibliography}
\end{document}